# Magnetization flip in Fe-Cr-Ga system

H. G. Zhang[1], B.T Song[1], J. Chen[1,2], M. Yue[1, *], E. K. Liu[2], W. H. Wang[2], G. H. Wu[2]

1 College of Materials Science and Engineering, Key Laboratory of Advanced Functional Materials, Ministry of Education of China, Beijing University of Technology, Beijing 100124, China

2State Key Laboratory of Magnetism, Institute of Physics, Chinese Academy of Sciences, Beijing 100190, China

**Abstract**

A systematic investigation about the structure and magnetism of $Fe_{75-x}Cr_{25}Ga_x$ (11<x<33) and $Fe_{50}Cr_{50-y}Ga_y$ (0<y<33) series has been carried out in this work. It shows that the parent $Fe_{50}Cr_{25}Ga_{25}$ phase has higher tolerance for Ga replacing Cr than replacing Fe atoms. An abrupt flip of Curie temperature and magnetization in the $Fe_{50}Cr_{50-y}Ga_y$ (0<y<33) series was observed at the composition of $Fe_{50}Cr_{25}Ga_{25}$. We proposed an explanation concerning anti-sites occupation and magnetic structure transition in this series. The induced structure is proved energetically favorable from first-principles calculations. This work can help us to understand the dependences between the crystal structure and magnetism in Fe-based Heusler compounds, and provides a method to deduce the atomic configurations based on the evolution of magnetism.

## 1. Introduction

The abundant physical phenomena and application possibilities of Heusler compounds are highly related to the atomic configurations of the elements in $X_2YZ$ (X and Y stand for the transition elements, Z represents main group elements and sometimes transition elements) [1-8]. Especially in Fe-based Heusler compounds, which has been theoretically predicted to be half-metallic ferromagnets or rare-earth-free hard magnets (in their tetragonal structure) [9-12]. However, most of these properties are impeded by the anti-sites disordering happened commonly in Fe-based Heusler compounds[13]. Though the reason for these disordering is not fully understood yet, one crucial factor we know is the magnetic interactions between different magnetic atoms. J. Kiss *et al.* [13] addressed that magnetic interactions between Fe atoms can affect the atomic ordering in $Fe_2CuGa$ and thus its magnetic properties. Our previous works in $Fe_2CrGa$ and $Fe_2CrAl$ systems also revealed the influences of magnetism on the atomic configurations[14]. Therefore, despite the conventional routine of determining the atomic configurations first and then interpret the magnetic behaviors, it is also possible to execute in a reversal way under some circumstances, i.e. deducing the atomic configurations based on the magnetic behaviors of particular Heusler alloys.

This idea is also valuable considering the difficulties one may encounter during the determination of atomic configurations in Heusler alloys. Conventional X-ray diffraction (XRD) method is not applicable in distinguishing the adjacent 3*d* transition elements (X & Y in Heusler alloys), owning to the small difference of their atomic scattering factors. Neutron diffraction is an ideal method which can provide a precise description about the distribution of the transition elements due to their distinguishable nuclear cross sections. It also can reveal the magnetic structure of the system[15, 16]. However, neutron diffraction spectrometer is not a common resource accessible for everyone in day-to-day laboratory work. Therefore, a simple solution that is convenient and effective is necessary for us to infer the atomic configurations in Heusler compounds.

Here we present an example about using element substitution and magnetic revolution to deduce the atomic configurations and magnetic structures in Fe-Cr-Ga system. A magnetization flip phenomenon was observed in the series. By analysing the variation of magnetizations together with different structure and magnetic models, we proposed an explanation including anti-sites occupation and magnetic structure transition for this phenomenon.

## 2. Material and methods

It is necessary to address the ideal atomic configurations of Heusler compounds before proceeding. As shown in Fig.1, a fully ordered Heusler compound usually crystallize in a $L2_1$ ($Cu_2MnAl$) or inverse $L2_1$ (X or $Hg_2CuTi$) structure, depending on which one of X and Y elements has the larger valence electron number [17-19]. The one with larger number of valence electron usually occupy the A (0, 0, 0) and C (0.5, 0.5, 0.5) sites, while the other one take the B (0.25, 0.25, 0.25) sites that is on the same plane with the main-group elements at D (0.75, 0.75, 0.75) sites. Any deviation from the two ideal structures will lead to a change of the magnetic behavior. Therefore, by replacing the three atoms in $Fe_2CrGa$ with each other, the evolution of magnetization will depend on which site the dopant goes. We thus chose the $Fe_{2-x}CrGa_x$ and $Fe_2Cr_{1-y}Ga_y$ series for the investigation.

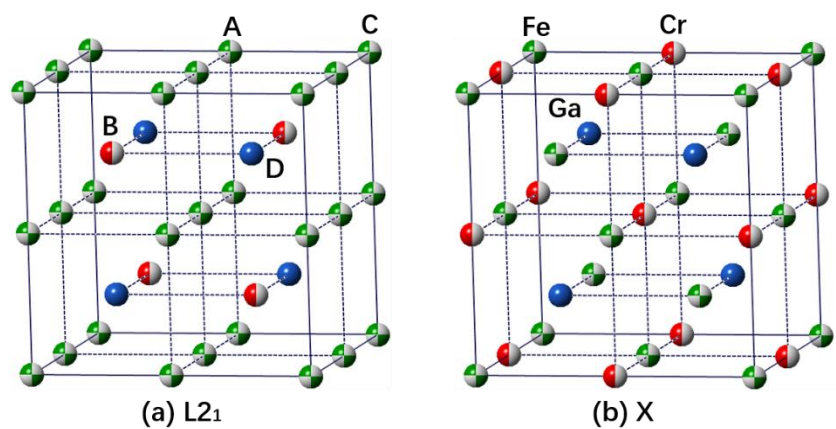

Fig. 1 (a) In $L2_1$ structure, the equivalent A, C sublattices are occupied by Fe (X) atoms, while B and D sublattices are occupied by Cr (Y) and Ga (Z) atoms, respectively; (b) In $Hg_2CuTi$ structure , Fe (X) atoms take A and B sites while Cr (Y) atom takes C sites.

$Fe_{75-x}Cr_{25}Ga_x$ (11<x<33) and $Fe_{50}Cr_{50-y}Ga_y$ (0<y<33) ingots were prepared by arc-melted method in an argon atmosphere and then homogenized at 1273 K for three days. X-ray diffraction (XRD) was performed in the region of 2θ=20-90° using Cu $K\alpha$ (λ=0.15406 nm) radiation (Rigaku Co., RINT-2400). Magnetization measurements were carried out using a homemade vibrating sample magnetometer (VSM) and superconducting quantum interference device (SQUID, Quantum Design MPMS-7).

For the theoretical calculations of the magnetic moment and total energy, we have performed first-principles calculation using AKAI-CPA-KKR (Korringa–Kohn–Rostoker Green's function method with CPA and LDA of the density functional method) codes based on the coherent potential approximation (CPA) [14, 20-22].

## 3. Results and discussion

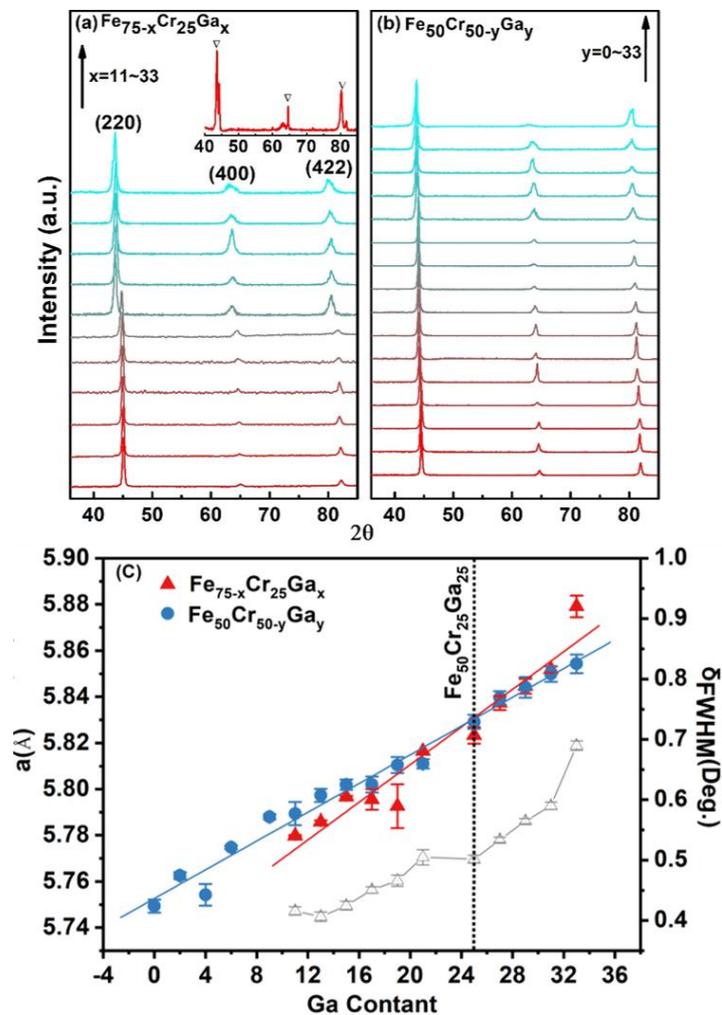

Fig.2. XRD patterns of the (a) $Fe_{75-x}Cr_{25}Ga_x$ (11<x<33) and (b) $Fe_{50}Cr_{50-y}Ga_y$ (0<y<33) series, where the inset shows the patterns of $Fe_{75-x}Cr_{25}Ga_x$ (x=35) samples; (c) shows the change of lattice parameters and $\delta_{FWHM}$ of the tow series along with the increase of Ga content.

The structure evolutions of the two series, $Fe_{75-x}Cr_{25}Ga_x$ (11<x<33) and $Fe_{50}Cr_{50-y}Ga_y$ (0<y<33), are shown in Fig.2. The Bragg diffraction peaks become broad along the increase of Ga atom in both series, as shown by the full width at half maximum ($\delta_{FWHM}$) in Fig.2(c). This indicates the increase of inner strain in the crystal lattice, since the contribution from size effect is small for these casted samples. The strain should come from the lattice distortion caused by radius difference between Fe, Cr and Ga atoms. It eventually leads to a collapse of the system and causes the split of the peaks for the sample with x=35 in $Fe_{75-x}Cr_{25}Ga_x$ system, as can be seen from the inset of Fig.2(a). The two sets of peaks suggest two phases with same symmetry and similar lattice parameters. The limit of doping in $Fe_{50}Cr_{50-y}Ga_y$ series is higher than that of the $Fe_{75-x}Cr_{25}Ga_x$ series, suggesting the $Fe_2CrGa$ phase has better tolerance for Ga replacing Cr than replacing Fe. This could be explained by that there is a continuous solid solution between Fe and Cr[23], and many compounds between Fe and Ga[24], while only a few compounds can be formed between Cr and Ga[25]. Ga substitution in both series leads to a monotonous increase of the lattice parameter, as shown in Fig.2(c). The only difference is that the lattice expand more rapidly when Ga replacing Fe than that of Ga replacing Cr. This can be explained by the larger size difference between Fe (1.16 Å) and Ga (1.24 Å) that that between Cr (1.22 Å) and Ga.

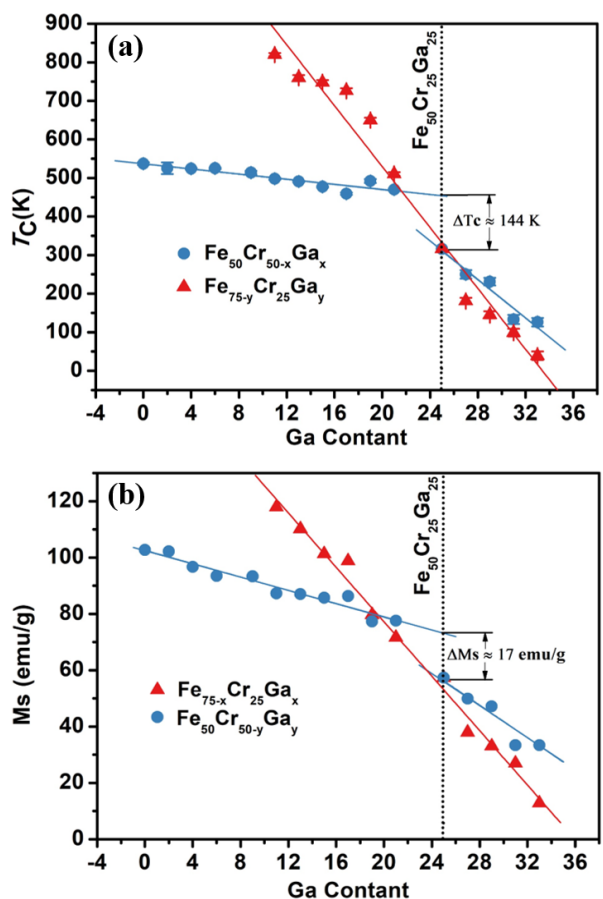

Fig.3 Ga content dependence of (a) Curie temperature and (b) saturation magnetization in $Fe_{75-x}Cr_{25}Ga_x$ (11<x<33) and $Fe_{50}Cr_{50-y}Ga_y$ (0<y<33) series. The symbols are the experimental data and the solid lines are the fitting curves. The dash line represents the stoichiometric composition of $Fe_{50}Cr_{25}Ga_{25}$.

The magnetic properties of the two series are shown in Fig.3. The Curie temperature ($T_C$) monotonously decreases along the increase of Ga content in the $Fe_{75-x}Cr_{25}Ga_x$ series. It suggests the weakening of ferromagnetic interactions between the Fe atoms due to the breakdown of the Fe sublattice. The $Fe_{50}Cr_{50-y}Ga_y$ series show a similar trend of $T_C$ but with an abrupt change as large as 144 K at the stoichiometric composition ($Fe_{50}Cr_{25}Ga_{25}$, i.e. $Fe_2CrGa$), as shown in Fig.3 (a). This abnormal flip cannot be simply explained by a sudden change of the interaction strength, since there is no drastic change of the crystal structures. Similar behaviors can also be observed for the magnetization variation in Fig.3 (b). The saturation magnetization ($M_S$) of the $Fe_{75-x}Cr_{25}Ga_x$ series shows an almost linear decrease with the increasing of Ga, while the $Fe_{50}Cr_{50-y}Ga_y$ series exhibit a flip at the composition of $Fe_{50}Cr_{25}Ga_{25}$. The tendency in the $Fe_{75-x}Cr_{25}Ga_x$ series is easy to explain: the replacement of Fe will always lead to a decrease of $M_S$ since Fe is the main contributor to the ferromagnetism. As for the flip in $Fe_{50}Cr_{50-y}Ga_y$ series, we need to do a systematic analysis about the variation of magnetization.

The substitution of Ga for Cr in $Fe_{50}Cr_{50-y}Ga_y$ series resembles a continuous transition of the composition from binary Fe-Cr to Fe-Ga alloy. Therefore, it is necessary to illustrate the magnetic and structural characters of these two end alloys. The Fe-Cr alloy has a very sensitive magnetic ground state which could be either ferromagnetic (FM) or anti-ferromagnetic (AFM), depending on the volume of the lattice[26-30]. At the equilibrium lattice volume, the ground state is FM, in which Fe and Cr moments are paralleled with each other, as shown in Fig.4 (a). When the lattice constant is increased for about 3%, an AFM structure becomes the ground state, with the Fe and Cr sublattices are respective AFM, as shown in Fig.4 (b). The Fe-Ga system is simply a FM A2-type structure with Fe and Ga equally occupied all the sites, as shown in Fig.4 (c)[30].

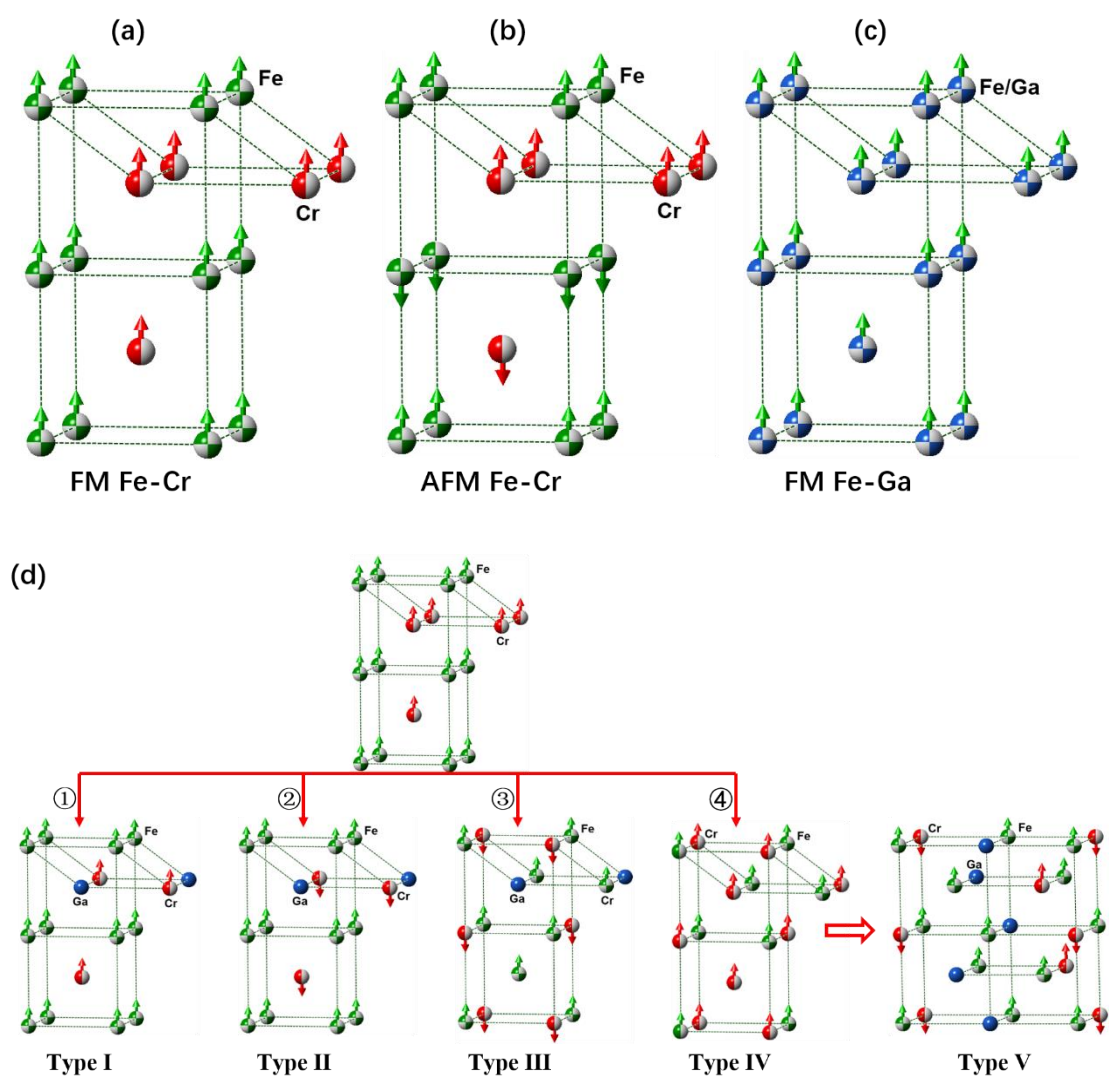

Fig.4 The crystal and magnetic structures of (a) FM Fe-Cr system, (b) AFM Fe-Cr system and (c) FM Fe-Ga system. Each model represents two molecules (FeCr or FeGa), i.e. four atoms, while the additional two atoms at right side of the unit cells are just for a better viewing of the structures. (d) The four scenarios related to the Ga substitution of Cr from Fe-Cr to Fe-Ga systems. Type I (①) and II (②) represent the FM and AFM configurations in the $L2_1$ structure of $Fe_{50}Cr_{25}Ga_{25}$, respectively. Type III (③) represents the AFM configuration in the $Hg_2CuTi$

structure of $Fe_{50}Cr_{25}Ga_{25}$. Type IV and V (④) represent the FM configuration in the B32 structure of Fe-Cr system and AFM configuration in the proposed structure of $Fe_{50}Cr_{25}Ga_{25}$ system, respectively.

There are many possible structures for the intermediate systems when replacing Cr with Ga in the $Fe_{50}Cr_{50-y}Ga_y$ series. First, we need to rule out those inconsequent ones based on the variation of magnetization. No matter it is the ground state or not, the AFM structure in Fig.4 (b) can be excluded immediately because replacing Cr atoms in this structure will lead to an increase of the overall magnetization. Therefore, we only need to consider the FM structure, and thereby obtain the following scenarios for Ga substitution.

If Ga replacing Cr as the way in scenario ①, the magnetization will decrease and result in a type-I structure at the composition of $Fe_{50}Cr_{25}Ga_{25}$. However, since the Cr atoms on both C & D sites possess the same magnetic moments, there will not be a flip of $M_S$ [28, 29]. Besides, this kind of replacement is not in favor of the A2 structure of Fe-Ga system. It is thus reasonable to infer that the flip of $M_S$ at $Fe_{50}Cr_{25}Ga_{25}$ should correspond to a magnetic reverse from FM to AFM. If the Cr atoms in type I structure reverse to AFM with respect to the Fe atoms, as shown in scenario ②, it will result in a type-II structure for $Fe_{50}Cr_{25}Ga_{25}$ which seems can fulfill the condition of $M_S$ flip. However, further replacement of Cr by Ga in type-II structure will lead to an increase of $M_S$, which is in contrast with the experimental result in Fig.3 (b).

The above discussion demonstrates that a reversal of magnetic moment is not sufficient to cause the $M_S$ flip. A structure variation has to be considered at this point, which is reasonable because the Fe-based Heusler compounds always have certain kind of anti-site disordering[13]. We thus propose a solution with structural transition caused by the anti-site disordering process in the $Fe_{50}Cr_{50-y}Ga_y$ series, as shown in scenario ③ and ④. One possible transition is from the $L2_1$-type to the $Hg_2CuTi$-type structure of $Fe_{50}Cr_{25}Ga_{25}$, as shown in scenario ③ with type-III structure. This transition is common but not suitable because it will lead to an increase instead of decrease of $M_S$ when Ga content is larger than 25, which is conflict with the experimental result. The other possible transition is from CsCl-type B2 structure to NaTl-type B32 structure of Fe-Cr system, as shown in scenario ④ with type-IV structure. The difference between the B2 and B32 structures is that the latter one shows equal occupation of both Fe and Cr atoms in the adjacent layers. According to the calculation, Fe or Cr atoms in adjacent lays possess the same moment in FM state [28, 29], Ga replacing Cr in adjacent layers thus will lead to a decrease of $M_S$ as shown in Fig.3 (b). Then, when Ga content reaches y=25, there will be a FM to AFM transition, caused by the moment reverse of the Cr atoms in the adjacent layers. This transition will lead to a sudden decrease (flip) of the $M_S$, because half of the FM contributions from Cr atoms suddenly become AFM contributions. The resulted type-V structure is a variant of the $Hg_2CuTi$ structure of Heusler alloys[31, 32]. Further replacement of Cr above y>25 range in this type-V structure will lead to a continuous decrease of $M_S$, in accordance with the experimental result in Fig.3 (b). In addition, since Ga atoms occupy the adjacent layers in this structure, it will naturally lead to an A2-type structure at the end composition of Fe-Ga alloy.

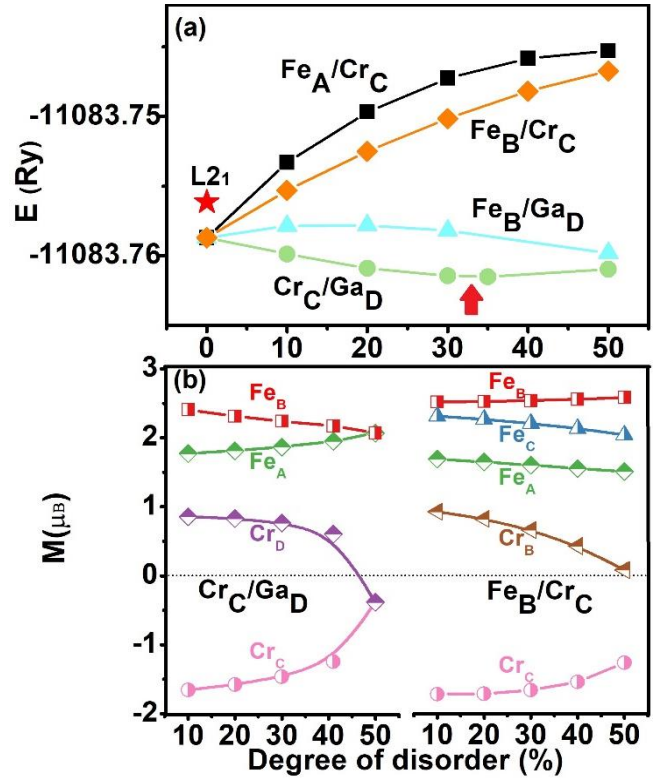

Fig.5 The calculated (a) total energies and (b) atomic moments of different kinds of anti-sites occupations (i.e. the $Fe_A$-$Cr_C$, $Fe_B$-$Cr_C$, $Fe_B$-$Ga_D$ and $Cr_C$-$Ga_D$ disordering) based on $Hg_2CuTi$ structure. The total energy of $L2_1$ structure is marked by the red star, while the stable state is marked by the red arrow in (a).

We then proved the validity of the second scenario by evaluating the stability of the derived type-V structure. A series of first-principles calculations were carried out for the related structures and the results are shown in Fig.5. The first conclusion is that the energy of $Hg_2CuTi$ structure is lower that of the $L2_1$ structure in $Fe_{50}Cr_{25}Ga_{25}$. Therefore, we just need to focus on the energy variations of $Hg_2CuTi$ structure with different kinds of anti-sites occupation, i.e. the $Fe_A$-$Cr_C$, $Fe_B$-$Cr_C$, $Fe_B$-$Ga_D$ and $Cr_C$-$Ga_D$ disordering as shown in Fig.5(a). There is only one stable state (indicated by the red arrow in Fig.5 (a)) can be obtained when the anti-sites occupation occurred between $Cr_C$ and $Ga_D$ atoms. It possesses a total energy lower than $Hg_2CuTi$ structure, while all the others are energetically unfavorable. This stable structure is perfectly agree with the derived type-V structure in Fig.4. Therefore, the proposed structure transition is reasonable in the Ga doped Fe-Cr system. It also reveals an interesting phenomenon that Ga element has a larger proportion on the D site (~70%) that on the C site (~30%). This is consistent with the model in Fig.4, which is crucial for the magnetization decrease when y>25.

The validity of scenario ② also be proved by the variation of magnetism shown in Fig.5(b). The dependences of atomic moments on the degree of disordering for $Cr_C$-$Ga_D$ and $Fe_B$-$Cr_C$ anti-sites occupation was compared, because only these two situation possess Cr atoms on the adjacent layers. One can see that, if Cr atoms appeared in adjacent layers, the $Cr_C$ atoms will be AFM with respect to the Fe atoms, while $Cr_D$ ones will remain in FM state. This is well agree with the FM to AFM transition we proposed in scenario ②. It also suggests that further substitution of Ga for Cr will lead to the decrease of magnetization as we discussed in Fig.4. Therefore, combining the conclusions based on Fig.5(a) and Fig.5(b), the proposed scenario ② is nicely qualified for the explanation of the observe magnetization flip.

**4. Conclusion**

In summary, we carried out a systematic investigation on the $Fe_{75-x}Cr_{25}Ga_x$ (11<x<33) and $Fe_{50}Cr_{50-y}Ga_y$ (0<y<33) series. It shows that the parent $Fe_{50}Cr_{25}Ga_{25}$ phase has higher tolerance for Ga replacing Cr than it replacing Fe atoms. Both substitution leads to a monotonous increase of the lattice parameter. However, the $T_C$ and $M_S$ of the $Fe_{50}Cr_{50-y}Ga_y$ (0<y<33) series shows a sudden flip at the composition of $Fe_{50}Cr_{25}Ga_{25}$. We thus made a thorough analysis about the possible situations of Ga substitution based on the magnetic evolution and the properties of Fe-Ga and Fe-Cr alloys. The result suggests that the flip of $T_C$ and $M_S$ can only be realized when the anti-sites occupation and a FM to AFM transition occurred simultaneously at $Fe_{50}Cr_{25}Ga_{25}$. The induced structure is based on the $Hg_2CuTi$ configuration with Ga atoms occupied both C and D sites. We further prove that this structure is energetically favorable compared with other $Hg_2CuTi$-based structures by using first-principles calculations. Its magnetic structure is also agree well with the experimental results. This work can help us to understand the relations between crystal structure and magnetism in Heusler compounds, and provides an important solution to deduce the atomic configurations by the evolution of magnetism.

**Acknowledgement:**

This work was supported by the Beijing Natural Science Foundation (No. 2182006), the General Program of Science and Technology Development Project of Beijing Municipal Education Commission of China (No. KM201710005006) and the National Natural Science Foundation of China [No.51401002].